\documentclass[preprint, preprintnumbers,amsmath,amssymb]{revtex4}
\usepackage{graphicx}% Include figure files
\usepackage{dcolumn}% Align table columns on decimal point
\usepackage{bm}% bold math

\def\be{\begin{equation}}
\def\ee{\end{equation}}
\def\bea{\begin{eqnarray}}
\def\eea{\end{eqnarray}}

\begin{document}

\markboth{M.Loewe, S. Mendizabal and R.A. Santos} {Gauge dependence
ambiguity and chemical potential in thermal U(1) theory}

\title{Gauge dependence ambiguity and chemical potential in thermal U(1) theory}

\author{\footnotesize M. Loewe\footnote{\email{mloewe@fis.puc.cl}}}
\author{S. Mendizabal \footnote{\email{smendiza@fis.puc.cl}}}
\author{R. A. Santos \footnote{\email{rasantos@puc.cl}}}

\address{Facultad de F\'\i sica,
Pontificia Universidad Cat\'olica de Chile,\\ Casilla 306, Santiago
22, Chile.}

\begin{abstract}
In this letter we explore the dependence on the gauge fixing
condition of several quantities in the $U(1)$ Higgs model at finite
temperature and chemical potential. We compute the effective
potential at the one loop level, using a gauge fixing condition that
depends on $\mu$ and which allows to decouple the contributions of
the different fields in the model. In this way we get the mass
spectrum and the characterization of the phase transition, pointing
out in each case how these quantities depend on the gauge fixing
parameter $\xi$. When $\mu$ vanishes, we agree with previous results
if $\xi=0$. The gauge dependence problem is also analyzed from the
perspective of the Nielsen identities.

\end{abstract}

\maketitle

\section{Introduction}

In spite of being a very simple case, the $U(1)$ Higgs model
acquires a quite complicated structure when we go to the scenario of
finite temperature and chemical potential $\mu$. For example, since
the paper by Bernard \cite{bernard}, it is well known that at finite
temperature the ghost fields cannot be factorized in a trivial way
from the functional integral in the $U(1)$ case, as it does occurs
at zero temperature.

The gauge fixing condition at finite temperature is a long standing
problem in field theory. Dolan and Jackiw \cite{dolan} explored the
gauge dependence of several quantities, in particular the critical
temperature from the effective potential. Keeping in mind the
validity of the one loop calculation up to order $1/\beta^2$, it is
possible to restore the gauge invariance by expanding the effective
potential up to this order.

The relationship between the gauge fixing condition and the
temperature for the phase transition in the electroweak theory has
been explored in \cite{ferrer1}. The analysis of the phase
transition at finite T, by the Landau method has been discussed in
\cite{ferrer2}. In particular, in the last article, the authors
found a critical temperature that does not depend on the gauge
parameter. However, this is due to the fact that a gauge condition
has been chosen, which couples the Higgs with the gauge field. As we
will see in this article, if we decouple the Higgs from the gauge
field, which is another possibility for fixing the gauge condition,
the critical temperature still depends on the gauge parameter $\xi$.

In the literature, people usually chooses the $\xi=1$ Feynman gauge,
or $\xi=0$ Landau gauge, claiming convenience reasons. When
discussing the effective potential in the frame of the
Weinberg-Salam model, the unitary gauge is excluded because it leads
to an erroneous result for the critical temperature and the pressure
of the system \cite{kapusta}. It seems that as soon as temperature
is turned on, the gauge invariance of the theory becomes a subtle
subject. In \cite{dittrich}, the authors emphasize that only
periodic gauge functions ($\Lambda_p$) are allowed, i.e. those that
satisfy $\Lambda_p(x^{\mu}+i\beta u^{\mu})=\Lambda_p(x^{\mu})$. In
fact, the gauge fixing procedure and the gauge invariance of
theories at finite temperature and density has not yet been fully
clarified. In particular, if we expand the effective potential up to
the order $1/\beta^2$ following \cite{dolan}, the gauge dependence
is not removed when finite chemical potential is present.

In this letter, we will consider the $U(1)$ Higgs model in the
presence of finite temperature and chemical potential. First, we
calculate the effective potential at the one loop level, using a
gauge fixing condition that depends explicitly on the chemical
potential, and which was used previously in a different context
\cite{lmr1, lmr2}. In this way, we get non trivial dispersion
relations for the different particles of our model. The role of the
ghost fields is analyzed in detail for different cases. We compare
our results, when $\mu=0$, with previous calculations
\cite{kapusta}, been in agreement with the results by Kapusta when
$\xi\rightarrow 0$. We calculate the critical temperature $T_c$
associated to the restoration of the $U(1)$ symmetry, emphasizing
its gauge dependence. The relationship between the chemical
potential and the gauge invariance problem is also analyzed.

%%%%%%%%%%%%%%%%%%%%%%%%%%%%%%%%%%%%%%%%%%%%%%%%%%%%%%%%%%%%%%%%%%%%%%

\section{The $U(1)$ Higgs Model}

Our model is described by the following Lagrangian

\begin{equation}\label{LagrangianoT0}
    \mathcal{L}=(D_\mu\phi)^*
D^\mu\phi-m^2(\phi^*\phi)-\lambda(\phi^*\phi)^2-\frac{1}{4}F^{\mu
\nu}F_{\mu \nu},
\end{equation}

\noindent where

\begin{eqnarray}
  D_\rho\phi &=&(\partial_\rho + ie A_\rho)\phi,  \\
  F_{\mu \nu} &=&\partial_\mu A_\nu - \partial_\nu A_\mu.
\end{eqnarray}

\noindent The set $\{\phi,\phi^*\}$, represents a pair of complex
scalar fields. We will start in the broken phase ($m^2<0$), where we
have a real non-vanishing vacuum expectation value $\nu$ for the
$\phi$ field. The local gauge transformations $U$ that leave our
Lagrangian invariant are the following

\begin{eqnarray}\label{A_mu_trans}
(A^{\mu}(x))^U &=&
A^{\mu}(x)+\partial_{x}^{\mu}\Lambda(x),\\
\tilde{\phi}^{U}&=&\phi^{U}-\nu,\nonumber\\
&=& \phi(x')-ie\Lambda(x')\phi(x')-\nu.\nonumber\
\end{eqnarray}

The partition function is given by

\begin{equation}\label{FNparti}
    Z=N(\beta)\int_{\mbox{Per.}}D\phi D\phi^* \prod_\rho\bar{DA}_\rho\exp{\int_0^\beta d\tau\int
    d^3\mathbf{x}\mathcal{L}_{eff}},
\end{equation}

\noindent where \emph{Per} in the functional integral indicates that
we have to integrate over periodic euclidean fields configuration in
the interval $(0,\beta)$.

The Lagrangian  $\mathcal{L}_{eff}$ in the previous equation,
incorporates the chemical potential through the recipe of
considering $\mu$ as the zero component of a constant external gauge
field \cite{actor},

\begin{eqnarray}
\nonumber\mathcal{L}_{eff}&=&
-(\partial_\rho-ieA_\rho+i\mu\delta_{\rho 0})\phi^*
(\partial_\rho+ieA_\rho-i\mu\delta_{\rho 0})\phi\\
&-&m^2\phi^*\phi-\lambda(\phi^*\phi)^2-\frac{1}{4}F_{\mu\nu}F_{\mu\nu}.
\end{eqnarray}

\section{One loop effective potential}

In order to calculate the effective potential, it is more convenient
to express the complex $\phi$ field as a linear combination of real
fields $\phi_1$ and $\phi_2$, such that
$\phi=\frac{1}{\sqrt{2}}(\phi_1+i\phi_2)$. $\phi_1$ acquires the non
vanishing expectation value $\nu$.

%We will The generating functional $W$ can be written as

%\begin{equation}\label{1loopaprox}
%    W\equiv-\ln Z\simeq-\mathcal{S}^{(0)}-\ln\left[\int_{\mbox{Per.}}D\phi_1 D\phi_2
%    DA\exp{(\mathcal{S}^{(1)}+\mathcal{S}^{(2)})}\right].
%\end{equation}
%ppppppppppppppppppppppppppppppppppppppppppppppppppppppppppppppppppppppppppppppppppppppppp

We will expand the action up to second order in powers of the
fields, using a Lagrangian $\mathcal{L}_J$ that incorporates
external sources, $J_1$, $J_2$ and $J_\mu$,

\begin{equation}
\mathcal{L}_{J}=\mathcal{L}_{eff}+\frac{J_1}{\beta
\Omega}\phi_1+\frac{J_2}{\beta \Omega}\phi_2+\frac{J_\rho}{\beta
\Omega}A_\rho.
\end{equation}

In this way we get

\begin{equation}\label{expS}
    \mathcal{S}=\int_0^\beta d\tau\int
    d^3\mathbf{x}\mathcal{L}_{T}
    =\mathcal{S}^{(0)}+\mathcal{S}^{(1)}+\mathcal{S}^{(2)}+...,
\end{equation}

\noindent where

\begin{equation}\label{S0}
    \mathcal{S}^{(0)}=\left[\frac{\mu^2}{2}\nu^2-
    \frac{m^2}{2}\nu^2-\frac{\lambda}{4}\nu^4\right]\beta\Omega+
    J_1\nu,
\end{equation}

\noindent and

\begin{eqnarray}\label{S1}
\nonumber\mathcal{S}^{(1)}&=&\int_0^\beta d\tau\int
    d^3\mathbf{x}\{-i\mu\nu\delta_{\rho 0}\partial_\rho\phi_2+\mu^2\nu\phi_1-
    eA_\rho\delta_{\rho 0}i\mu\nu^2-m^2\nu\phi_1 -\lambda\nu^3\phi_1\\
    &+& \frac{J_1}{\beta\Omega}\phi_1
    +\frac{J_2}{\beta\Omega}\phi_2+\frac{J_\rho}{\beta\Omega}A_\rho\}.
\end{eqnarray}

By shifting $\phi_1 \rightarrow \phi_1 + \nu$, we can suppress terms
proportional to $\phi_1$ in $\mathcal{S}^{(1)}$, since the quantum
fields must have a zero vacuum expectation value. Choosing

\begin{equation}%\label{}
\frac{J_{\phi_1}}{\beta\Omega}=\nu(m^2-\mu^2+\lambda\nu^2),
\end{equation}

\noindent and

\begin{equation}%\label{}
    \frac{J_{A_0}}{\beta\Omega}=ei\mu\nu^2,
\end{equation}

\noindent we can eliminate the terms proportional to $\phi_1$ and
$A^0$ in $\mathcal{S}^{(1)}$.

The $\mathcal{S}^{(2)}$ can be written as a quadratic form

\begin{eqnarray}\label{S2explicit}
    \mathcal{S}^{(2)}=-\frac{1}{2}\int_0^\beta d\tau\int
    d^3\mathbf{x}(\phi_1, \phi_2, A_\rho)\mathbb{M}\left( \begin{array}{c} \phi_1 \\ \phi_2 \\
    A_\alpha\end{array}\right),
\end{eqnarray}

\noindent where $\mathbb{M}$ is a matrix given by \small
\begin{eqnarray}
\mathbb{M}=\left(%
\begin{array}{ccc}
  -\partial_\rho\partial_\rho-\mu^2+m^2+3\lambda\nu^2 & 2i\mu\partial_\rho\delta_{\rho 0} & 2i\mu\delta_{0 \rho}\nu \\
  -2i\mu\partial_\rho\delta_{\rho 0} & -\partial_\rho\partial_\rho-\mu^2+m^2+\lambda\nu^2 & e\nu\partial_\rho \\
  2i\mu\delta_{0 \alpha}\nu & -e\nu\partial_\alpha & e^2\delta_{\rho\alpha}\nu^2+ B_{\rho\alpha} \\
\end{array}%
\right).\quad
\end{eqnarray}
\normalsize

The operator $B_{\rho\alpha}$ is defined as

\begin{equation}\label{Bmunu}
    B_{\rho\alpha}=-\partial_\lambda\partial_\lambda\delta_{\rho
    \alpha}+ \partial_\rho\partial_\alpha.
\end{equation}

In the previous matrix we have undesirable mixing terms which
couples $\phi$ with A. These terms can be eliminated through an
appropriate gauge fixing condition given by

\begin{equation}\label{fijacionG}
    F=(\partial_\rho+i2\mu\delta_{\rho
    0})A_\rho-ie\xi\nu(\phi_1+i\phi_2).
\end{equation}

This gauge fixing condition was introduced previously \cite{lmr1} to
calculate the contribution of each field to the the effective
potential of the Weinberg-Salam model, and as a background field
that maintains explicitly the gauge invariance \cite{lmr2}.

From this condition,  the Faddeev-Popov Lagrangian, at finite $\mu$
can be read as

\begin{equation}%\label{}
    \mathcal{L}_{FP}=\partial_\rho\bar{\eta}(\partial^{\rho}+2i\mu\delta_{\rho
    0})\eta-\xi \bar{\eta}e^2\nu^2\eta
\end{equation}

Finally, our total lagrangian is written as a sum of the different
contributions
$\mathcal{L_T}=\mathcal{L}_J+\mathcal{L}_{GF}+\mathcal{L}_{FP}$. The
ghost Lagrangian plays a important role when counting the degrees of
freedom of the model, but since they do not couple to the physical
fields, their role will be discussed in the next section. The action
term $\mathcal{S}^{(2)}$ is defined by a new matrix
$\mathbb{\tilde{M}}$,

\begin{eqnarray}
\mathbb{\tilde{M}}=\left(%
\begin{array}{ccc}
  -\partial_\rho\partial_\rho-\mu^2+ {m}_1^2 & 2i\mu\partial_\rho\delta_{\rho 0} & 0 \\
  -2i\mu\partial_\rho\delta_{\rho 0} & -\partial_\rho\partial_\rho-\mu^2+{m}_2^2& 0 \\
  0 & 0 & e^2\delta_{\rho\alpha}\nu^2+ \widetilde{B}_{\rho\alpha} \\
\end{array}%
\right),\quad
\end{eqnarray}

\noindent where we have introduced effective gauge dependent masses
 ${m}_1$, ${m}_2$ as

\begin{eqnarray}
  {m}_1^2 &=&  m^2+3\lambda\nu^2 +e^2\nu^2\xi,\\
  {m}_2^2 &=&  m^2+\lambda\nu^2 +e^2\nu^2\xi.
\end{eqnarray}

Note that both $m_1$ and $m_2$ depend on the gauge parameter $\xi$
this means, for an arbitrary value of $\xi$ there are no Goldstone
boson in this model. We will come back to this point later.

The operator $\widetilde{B}_{\rho\alpha}$ is the extension of
(\ref{Bmunu}) including gauge dependent terms

\begin{equation}\label{Btilde}
    \widetilde{B}_{\rho\alpha}=-\partial_\lambda\partial_\lambda\delta_{\rho
    \alpha}+ \partial_\rho\partial_\alpha(1-\xi^{-1})+4\mu^2\delta_{\rho 0}\delta_{\alpha
    0}\xi^{-1}.
\end{equation}

Finally, the effective potential is given by

%\begin{equation}%\label{}
%\mathcal{V}_{eff}=,
%\end{equation}

\begin{equation}%\label{}
\mathcal{V}_{eff}=-\frac{1}{\beta\Omega}(\ln
Z)=-\frac{1}{\beta\Omega}(\mathcal{S}^{(0)}+\ln Z_1),
\end{equation}

\noindent with

\begin{equation}
Z_1\equiv\int_{\mbox{Per.}}D\phi_1 D\phi_2
  DA e^{\mathcal{S}^{(2)}},
\end{equation}

\noindent where $\Omega$ is the spatial volume.

In the momentum space, $\ln Z_1$ can be written as a sum over the
Matsubara frequencies $\omega_n=2\pi n/\beta$

\small\noindent
\begin{eqnarray}
\nonumber &\ln Z_1=-\frac{\Omega}{2}\sum_n\int
\frac{d^3\mathbf{k}}{(2\pi)^3}\ln \left\{\det_{\rho\alpha}([k^2+
e^2\nu^2]\delta_{\rho\alpha}-k_\rho k_\alpha (1-\xi^{-1})-
4\mu^2\delta_{\rho 0}\delta_{\alpha 0}\xi^{-1})\right\}\\\nonumber
&-\frac{\Omega}{2}\sum_n\int \frac{d^3\mathbf{k}}{(2\pi)^3}\ln
\left\{k^4+k^2({m}_1^2+ {m}_2^2-2\mu^2)+4\mu^2\omega_n^2+
({m}_1^2-\mu^2)({m}_2^2-\mu^2)\right\},\quad \\
\qquad
\end{eqnarray}
\normalsize

\noindent with $k^2=\omega_n^2+\mathbf{k}^2$ and where
$\det_{\rho\alpha}$ refers to the determinant in the Lorentz
indices. The polynomials in the equation above, since they are
quadratic in $\omega_n^2$, can be factorized such that

\small\noindent
\begin{eqnarray}\label{LNz1}
\nonumber &\ln Z_1=-\frac{\Omega}{2}\sum_n\int
\frac{d^3\mathbf{k}}{(2\pi)^3}\ln
\{(\omega_n^2+x_1^2)(\omega_n^2+x_2^2)\}\\\nonumber
&-\frac{\Omega}{2}\sum_n\int \frac{d^3\mathbf{k}}{(2\pi)^3}\ln
\{(\omega_n^2+y_1^2)(\omega_n^2+y_2^2)\}+\ln{\{(\omega_n^2+ y_3^2)(\omega_n^2+y_4^2)}\},\quad \\
\end{eqnarray}
\normalsize

Using the identity

\begin{equation}\label{identity}
\sum_n\ln (\omega_n^2+X^2)=\beta X+2\ln(1-e^{-\beta X}),
\end{equation}

\noindent we can decompose the effective potential as a sum of terms
associated to the different fields in the model. In other words we
have diagonalized the effective potential. Then

\begin{equation}
V_{eff}=V_{\phi}+V_{A},
\end{equation}

\noindent where $V_{\phi,A}$ is given by

\begin{equation}%\label{}
    V =-\int\frac{d^3\mathbf{k}}{(2\pi)^3}\frac{X(k)}{2}+\frac{\ln(1-e^{-\beta
    X(k)})}{\beta},
\end{equation}

\noindent which is valid for each component of the fields.

In this way we can identify $x_{1,2}$ and $y_{i}$ $(i=1..4)$ with
the energy spectra of these pseudo-particles

\begin{eqnarray}\nonumber\label{pseudo}
x_1^2&=&m_1^2+\mathbf{k}^2-\lambda\nu^2+\mu^2+\sqrt{4\mu^2(m_1^2+\mathbf{k}^2-\lambda\nu^2)
+\lambda^2\nu^4},\qquad \\
x_2^2&=&m_2^2+\mathbf{k}^2+\lambda\nu^2+\mu^2-\sqrt{4\mu^2(m_2^2+\mathbf{k}^2+\lambda\nu^2)
+\lambda^2\nu^4}.\qquad
\end{eqnarray}

Notice that if $\nu=0$ and $\mu\neq0$, (\ref{pseudo}) simplifies to
the well known expression

\begin{equation}%\label{}
    x_{1,2}=E^\phi_\pm=\sqrt{\mathbf{k}^2+m^2}\pm\mu.
\end{equation}

\noindent Here, $m_1=m_2=m$ and there is no dependence on $\xi$. If
$\mu=0$ and $\nu\neq0$, $x_1$ and $x_2$ depend on $\xi$.

For the gauge fields we find two non-anomalous massive ``photonic"
states which obeys the following dispersion relation

\begin{equation}
y_1=y_2=E_\gamma=\sqrt{k^2+e^2\nu^2}.
\end{equation}

Notice that $x_1$ and $x_2$ describe the possible energy values for
$\phi_1$ and $\phi_2$.

The pathological energies associated to the other two ``photonic"
degrees of freedom (\ref{LNz1}) are given, in terms of
$\varepsilon=1-\xi$, by

\small
\begin{eqnarray}\label{y1_y2}\nonumber
y_3^2 \!\!\!&=&\!\!\!
{\mathbf{k}}^{2}+\frac{{e}^{2}{\nu}^{2}}{2}(1+\xi)+2{\mu}^{2}+\frac{1}{2}\sqrt
{{e}^{4}{\nu}^{4}\varepsilon^2+8{\mu}^{2}{e}^{2}{\nu}^{2}\varepsilon-16{\mu}^{2}({\mathbf{k}}^{2}\xi^{-1}\varepsilon-\mu^2)},
\qquad\\
y_4^2\!\!\!&=&\!\!\!{\mathbf{k}}^{2}+\frac{{e}^{2}{\nu}^{2}}{2}(1+\xi)+2{\mu}^{2}-\frac{1}{2}\sqrt
{{e}^{4}{\nu}^{4}\varepsilon^2+8{\mu}^{2}{e}^{2}{\nu}^{2}\varepsilon-16{\mu}^{2}({\mathbf{k}}^{2}\xi^{-1}\varepsilon-\mu^2)}.
\qquad
\end{eqnarray}
\normalsize

We call these energies pathologicals, since they do depend
explicitly on the gauge parameter $\xi$.

Repeating the same previous analysis, if $\nu=0$ and $\mu\neq0$,
with $\xi=1$, we recover a photon with energy $y_4=|k|$.

On the other side, in the broken phase when $\mu=0$, we have
$y_3=\sqrt{\mathbf{k}^2+e^2\nu^2}$ and
$y_4=\sqrt{\mathbf{k}^2+e^2\nu^2\xi}$, i.e. we have three massive
degrees of freedom ($y_1,y_2$ and $y_3$) for the photon and a gauge
dependent contribution ($y_4$). Now taking $\mu\neq0$ and $\xi=1$,
we recover the three degrees of freedom for the massive photon:
$y_1, y_2$ and $y_4$.

As we will see in the next section, the Faddeev-Popov ghost fields
will compensate these anomalous dispersion relations. In particular
for $\xi =1$, the situation is completely clear. This probably is
one of the reasons why people prefer the Feynman Gauge $\xi=1$.

\section{Ghosts and Effective Potential}

In order to get a proper counting of the degrees of freedom, we have
to introduce the Faddeev-Popov ghost fields. This point is
completely different with respect to what occurs in the U(1) theory
at zero temperature, where the ghost can be factorized and do not
contribute to the effective potential.

The functional integration over the ghost fields, represented by the
complex Grassmann variables $(\eta,\bar{\eta})$, leaves

\begin{equation}%\label{}
    \int\!\! D\eta D\bar{\eta}\exp\left\{\int_0^\beta\!\!\int d^3\mathbf{x}\bar{\eta}
    \left(-\frac{\delta F}{\delta
    \Lambda}\right)\eta\right\}=\Omega\sum\!\!\!\!\!\!\!\!\int\,
    \ln(\omega_n^2+\mathbf{k}^2+e^2\nu^2\xi+2i\mu\omega_n).
\end{equation}

If we write the argument of the logarithm in the polar
representation

\begin{equation}%\label{}
    \rho
    e^{i\varphi}=\omega_n^2+\mathbf{k}^2+e^2\nu^2\xi+2i\mu\omega_n,
\end{equation}

\noindent with

\begin{eqnarray}%\label{}
\noindent\rho&=&\sqrt{(\omega_n^2+\mathbf{k}^2+e^2\nu^2)^2+4\mu^2\omega_n^2},\\
\varphi&=&\arctan\left(\frac{4\omega_n\mu}{\omega^2+\mathbf{k}^2+e^2\nu^2}\right),
\end{eqnarray}

\noindent we see that the sum over $\varphi$ vanishes since
$\varphi(\omega_n)=-\varphi(-\omega_n)$. Factorizing
$\rho^2=(\omega_n^2+z_1^2)(\omega_n^2+z_2^2)$, we find

\begin{equation}
z_{1,2} =E^\eta_\pm =\sqrt{\mathbf{k}^2+e^2\nu^2\xi+\mu^2}\pm \mu. \\
\end{equation}

If $\mu=0$, it is interesting to notice that one of the ghost fields
cancels the contribution from the nonphysical photon $y_3$

Using \ref{identity}, we can decompose the effective potential such
that $V_{eff}=V_{tree}+V_{\phi}+V_{A}+V_{\eta}$, where the different
contributions are

\begin{itemize}
    \item Tree level contribution

    \begin{equation}%\label{}
V_{tree}=\left[\frac{m^2}{2}\nu^2-\frac{\mu^2}{2}\nu^2+\frac{\lambda}{4}\nu^4\right];
\end{equation}
    \item $\phi$ contribution

\begin{equation}%\label{}
    V_\phi =\int\frac{d^3\mathbf{k}}{(2\pi)^3}\frac{(x_1(k)+x_2(k))}{2}+\frac{\ln(1-e^{-\beta
    x_1(k)})(1-e^{-\beta x_2(k)})}{\beta};
\end{equation}

    \item Massive photons contribution
\begin{eqnarray}%\label{}
\nonumber
    V_A &=&\int\frac{d^3\mathbf{k}}{(2\pi)^3}\frac{(y_1(k)+y_2(k))}{2}+\frac{\ln(1-e^{-\beta
    y_1(k)})(1-e^{-\beta y_2(k)})}{\beta}\\
    &+&\sqrt{\mathbf{k}^2+e^2\nu^2}+\frac{2}{\beta}\ln(1-e^{-\beta\sqrt{\mathbf{k}^2+e^2\nu^2}});
\end{eqnarray}
    \item Ghost contribution
\begin{equation}%\label{}
    V_\eta =-\int\frac{d^3\mathbf{k}}{(2\pi)^3}\frac{(z_1(k)+z_2(k))}{2}+\frac{\ln(1-e^{-\beta
    z_1(k)})(1-e^{-\beta z_2(k)})}{\beta}.
\end{equation}
\end{itemize}

\section{High Temperature expansion}

Our previous results for the effective potential are interesting
since we decouple the contributions for the different fields.
However, the integrals cannot be calculated analytically, and
therefore it is appealing to carry on a high temperature expansion
of the effective potential, which will allow us to compare our
expressions with well known results from the literature.

For the one loop effective potential in a high temperature expansion
we find

\begin{equation}\label{potefffin}
    V_{eff}=\left[\frac{m^2}{2}\nu^2-\frac{\mu^2}{2}\nu^2+\frac{\lambda}{4}\nu^4\right]
-\frac{2\pi^2T^4}{45}+\frac{T^2}{12}\left\{m^2+2\lambda\nu^2+\frac{(3+\xi)e^2\nu^2}{2}+\xi^{-1}\mu^2\right\}.
\end{equation}

The vacuum is defined by

\begin{eqnarray}%\label{}
    &&\lambda\nu(T)_{min}^2=\left\{%
\begin{array}{ll}
    (\mu^2+|m|^2)\left[1-\frac{T^2}{T_c^2}\right], & \hbox{para $T\leq T_c$;} \\\label{nu(T)}
    0, & \hbox{para $T>T_c$.} \\
\end{array}%
\right.\\
   && T_c^2=\frac{12(\mu^2+|m|^2)}{4\lambda+3e^2+\xi
e^2}\label{TC},
\end{eqnarray}

\noindent where $T_c$ is the critical temperature where the symmetry
is
 restored.

 The critical reader at this moment could be surprised that the
 critical temperature, which is in principle an observable quantity,
 depends on the gauge parameter. Notice, however, that for finite
 chemical potential it is well known that the number of Goldstone
 bosons is lesser than the usual prediction (\cite{Sholkovy}).
 This is in agreement with our results.
 In our case, both higgs fields acquire a mass. On the other hand, if $\mu$ vanishes, we
 know  that we must recover one Goldstone boson. This is not
 possible, unless that $\xi$ also vanishes. In this case the $R_\xi$
 gauge becomes the Lorentz gauge. Our results is then in agreement with \cite{kapusta} and \cite{ferrer1},
when
 $\mu=0$, in the gauge defined as $\xi=0$ (Landau gauge). So this
 gauge seems to be the more appropriate to extend the calculations
 to the scenario with finite mu.

 For finite $\mu$, it seems that the one loop calculation of the
 effective potential is not enough for having a gauge independent
 result for the critical temperature.

 In general, from previous work \cite{Pisarski} we know that the
 determination of critical values of parameters associated to phase
 transitions requires to go beyond the one loop approximation
 through an appropriate re-summation.

\section{Symmetry restoration, phase transition and Nielsen Identities}

As it is well known, the occurrence of the phase transition can be
inferred from the behavior of the isothermals of the effective
potential. The pressure is defined as $P=T\frac{\partial}{\partial
V}\ln Z,$ which implies $P=-V_{eff}$, since the effective potential
corresponds to the thermodynamical potential of the system.

We found for the pressure.

\begin{eqnarray}%\label{}
&&\!\!\!\!\!\!\!\!\!\!\!\!\!P_<=\frac {2{\pi
}^{2}{T}^{4}}{45}+\frac{{|m|}^{2}-\mu^{2}\xi^{-1}}{12}{T}^{2}+
    \frac{({|m|}^{2}+{\mu}^{2})^2}{4\lambda}\!\left(1-{\frac {{T}^{2}}{{T_{{c}}}^{2}}}\right)^2,\\
&&\!\!\!\!\!\!\!\!\!\!\!\!\!P_>=\frac {2{\pi
}^{2}{T}^{4}}{45}+\frac{{|m|}^{2}-\mu^{2}\xi^{-1}}{12}{T}^{2},
\end{eqnarray}

\noindent where $P_<$ $(P_>)$ corresponds to the broken (symmetric)
phase. Here we have the same difficulty, a gauge dependence, we
found previously for the critical temperature.

The pressure and the entropy are continuous at $T_c$. However the
specific heat has a discontinuity which depends on $\xi$ only
through $T_c$. This means that this is a second order phase
transition. In fact,the behavior of the specific heat confirms this
picture.

%\begin{figure}
%\includegraphics[angle=0,width=0.5\textwidth]{Graph1.eps}
%\caption{\label{specheat} The specific heat as a function of
%Temperature with $\xi=1$, for different values of $\mu$. ($\mu=0$
%(MeV): dashed line; $\mu=100$ (MeV): solid line; $\mu=200$ (MeV):
%doted line.}
%\end{figure}

The gauge dependence problem can also be analyzed from the
perspective of the Nielsen Identities. This method \cite{Nielsen},
related to the BRST symmetry transformations, is a procedure that
allows to search for possible gauge dependence of physical
quantities which are related to an explicitly gauge dependent
effective action. If these identities are satisfied, we may have
confidence that the results for the physical magnitudes will be
gauge independent. The Nielsen identities are still valid for finite
temperature \cite{asdokdas}. An interesting discussion of the
Nielsen identities for the generalized $R_\xi$ gauge in the abelian
Higgs model can be found in \cite{lauta}. We will follow the
strategy of this article to explore the validity of the Nielsen
identities for our case.

The Nielsen identities arise from the following identity

\begin{equation}
\xi\frac{\partial\Gamma}{\partial\xi}=\int d^dx\int d^dy
\frac{\delta\Gamma}{\delta\phi_i(y)}\left\langle\Delta_i\eta(y)\bar{\eta}(x)\left[\frac{F}{2}-\xi\frac{\partial
F}{\partial\xi}\right]\right\rangle_\Gamma,
\end{equation}

where we use the notation

\begin{equation}
\langle O\rangle =e^\Gamma\int D\phi_i D\eta D\bar{\eta} O
\exp{\left(-S_F+\int d^dx \frac{\delta\Gamma}{\delta
\phi_i}(\phi_i-\varphi_i)\right)},
\end{equation}

 and where $\eta , \bar{\eta}$ are the ghost fields, $F$ is the gauge fixing
 condition, which in our case is given by eq. (\ref{fijacionG}), $\Gamma$
 is the effective action, $\xi$ is the gauge parameter and $\Delta_i$ is
 the BRST transformation of the scalar fields $\phi_1$ and $\phi_2$
 which is given by $\delta \phi_i$= $\Delta_i\eta$ with
 $\Delta_i=(-\phi_2,\phi_1)$.

 In the finite temperature scenario, we have to keep in mind that
 the integrals in the previous equation, when going to the momentum space must be handle (d=4) as
\begin{equation}
\int d^4k\rightarrow \sum_n\int \frac{d^3k}{(2\pi)^3},
\end{equation}

where we sum over Matsubara frecuencies $\omega_n=2\pi n/\beta$.

Assuming for the classical components of the fields $\phi_1,\phi_2$,
$\phi_1=\nu$ and $\phi_2=0$, the Nielsen identities can be expressed
as

\begin{equation}\label{nielsenid}
\xi\frac{\partial V}{\partial \xi}=C\frac{\partial V}{\partial \nu},
\end{equation}

where C is given by

\begin{equation}
C=\frac{1}{2}\int d^dx\int d^dy
\left\langle\phi_2(y)\eta(y)\bar{\eta}(x)\left[
(\partial_\rho+i2\mu\delta_{\rho
    0})A_\rho+ie\xi\nu(\phi_1+i\phi_2).\right]\right\rangle,
\end{equation}

\noindent and must be evaluated in a perturbative expansion, see
\cite{lauta}.

We explored the validity of the Nielsen identities for the following
cases

\begin{itemize}
  \item  $\mu=0, \nu=0$:
It is very easy to see that the equation (\ref{nielsenid}) is valid
for any value of $\xi$, in ths case, confirming \cite{asdokdas} the
validity of the Nielsen identities for finite temperature.

  \item  $\mu=0, \nu\neq0$ In this case the Nielsen identities are valid only if we
  take $\xi=0$ and if we evaluate the derivatives in $\nu=\nu_{min}$, i.e. in the classical value. Note
  that $\xi =0$ is demanded by the Goldstone theorem, which implies $m_2=0$
  \item $\mu\neq0, \nu=0$ and  $\mu\neq0, \nu\neq0$.  In these cases the Nielsen identities
  are not valid for any possible value of $\xi$. In fact, the
  only non vanishing contribution to the left hand side of the Nielsen
  identities, according to eq. (\ref{nielsenid}) is given by the ``photonic" dispersion
  relations, eq. (\ref{y1_y2}), for the $\mu=0, \nu\neq0$ case. When
  $\mu\neq0, \nu\neq0$, all degrees of freedom contribute to the
  derivatives that appear on the Nielsen identities. However, in
  both cases the $\frac{\partial y_i}{\partial \xi}$ diverge when $\xi$
  goes to zero. This means that we are not allowed to take the
  $\xi=0$ gauge, where the Goldstone theorem is realized, in
  agreement with the result by \cite{Sholkovy}.
\end{itemize}

To summarize we can see that the $\mu$ dependent gauge fixing
condition we used here, and which was valuable for diagonalizing the
effective potential in the Weinberg-Salam model for $\xi=1$,
\cite{lmr1}, turn out to be unnatural when looking for gauge
invariant predictions for physical quantities in the abelian Higgs
model.

The gauge dependence problem is related to two different aspects:

\begin{itemize}
  \item The necessity of having gauge transformation which are periodic in
the temporal direction, with period $\beta$ and
  \item the occurrence of finite $\mu$.
\end{itemize}

As a conclusion, we would like to remark that the problem of the
gauge fixing at finite temperature and/or density has not yet been
solved. In the U(1) Higgs model we have explored in detail how this
gauge dependence propagates through several physical relevant
quantities. Nevertheless, when $\mu$ vanishes, we agree with
previous results in the literature.

It can be shown that $\mu$ can be incorporated as a boundary
condition for the Green function of fields confined to a finite
spatial region, as it occurs in chiral bag models
\cite{falomir1,falomir2}. Eventually, the construction of the theory
should start by taking the theory in a bounded space region, with
adequate $\mu$-dependent boundary condition. When taking then the
limit where we go into the whole space, probably a remanent of the
non-trivial boundary conditions will remain as a non trivial
dependence on the gauge fixing condition. This point will be
explored in a future work.

\section*{ACKNOWLEDGMENTS}

Financial support from
 FONDECYT under grant 1051067 is acknowledged. The authors would like to thank Dr.
 E.J.
 Ferrer for private communication. The authors also thank Drs. L. Vergara
 and C. Contreras for educating us about the Nielsen identities in
 our problem. M.L acknowledges support from the Centro de Estudios
 Subat\'{o}micos.

\end{document}